\documentclass[twocolumn,10pt,cleanfoot]{asme2e}

\usepackage{graphicx}
\usepackage{amsmath}    
\usepackage{enumerate}
\usepackage{multirow}
\usepackage{color}
\usepackage{booktabs}
\usepackage{amsfonts}
\usepackage{algorithm}
\usepackage{algpseudocode}
\usepackage{setspace}
\usepackage[numbers]{natbib}
\usepackage{bm}
\usepackage{stackengine}

\usepackage{array}
\newcolumntype{L}[1]{>{\raggedright\let\newline\\\arraybackslash\hspace{0pt}}m{#1}}
\newcolumntype{C}[1]{>{\centering\let\newline\\\arraybackslash\hspace{0pt}}m{#1}}
\newcolumntype{R}[1]{>{\raggedleft\let\newline\\\arraybackslash\hspace{0pt}}m{#1}}

\newcommand{\indentStyle}[0]{\indent}

\title{Characterization of Model-Based Uncertainties in Incompressible Turbulent Flows by Machine Learning}

\author{Mustafa Usta\thanks{Address all correspondence to this author.}~\thanks{Author was with the Department of Mechanical Engineering at Lehigh University, Bethlehem, PA.}
	\affiliation{
		G. W. Woodruff School of Mechanical Engineering\\
		Georgia Institute of Technology\\
		Atlanta, GA 30332\\
		Email: mustafa.usta@me.gatech.edu 
	}	
}

\author{Ali Tosyali
	\affiliation{
		Department of Industrial and Systems Engineering\\
		Rutgers, The State University of New Jersey\\
		Piscataway, NJ 08854\\
		Email: ali.tosyali@rutgers.edu
	}
}

\begin{document}

\maketitle    

\begin{abstract}
	This work determines the inaccuracy of using Reynolds averaged Navier Stokes (RANS) turbulence models in transition to turbulent flow regimes by predicting the model-based discrepancies between RANS and large eddy simulation (LES) models. Then, it incorporates the capabilities of machine learning algorithms to characterize the discrepancies which are defined as a function of mean flow properties of RANS simulations. First, three-dimensional CFD simulations using k-omega Shear Stress Transport (SST) and dynamic one-equation subgrid-scale models are conducted in a wall-bounded channel containing a cylinder for RANS and LES, respectively, to identify the turbulent kinetic energy discrepancy. Second, several flow features such as viscosity ratio, wall-distance based Reynolds number, and vortex stretching are calculated from the mean flow properties of RANS. Then the discrepancy is regressed on these flow features using the Random Forests regression algorithm. Finally, the discrepancy of the test flow is predicted using the trained algorithm. The results reveal that a significant discrepancy exists between RANS and LES simulations, and ML algorithm successfully predicts the increased model uncertainties caused by the employment of k-omega SST turbulence model for transitional fluid flows.\\
\end{abstract}
%
\section*{NOMENCLATURE}
	\begin{table}[H]
        \small
		\raggedright
		\renewcommand{\arraystretch}{1.2}
		\vspace{-0.3in}
		\begin{tabular}{ll}
			$d$ & cylinder diameter $\mathrm{[m]}$ \\
			$F_1$ & blending function $\mathrm{[-]}$
		\end{tabular}
	\end{table}
	\begin{table}[H]
		\small
		\raggedright
		\renewcommand{\arraystretch}{1.2}
		\begin{tabular}{ll}
			$k$ & turbulent kinetic energy $\mathrm{[m^2/s^2]}$ \\
			$p$ & pressure $\mathrm{[Pa]}$ \\
			$Re$ & Reynolds number $\mathrm{[-]}$, $Re=Ud/\nu$ \\
			$t$ & time $\mathrm{[s]}$ \\
			$\vec{u}$ & flow velocity vector \\
			$U_a$ & average inlet velocity \\
			$\beta^*$ & turbulence closure coefficient $\mathrm{[-]}$ \\
			$\epsilon$ & dissipation rate $\mathrm{[m^2/s^3]}$ \\
			$\gamma$ & closure coefficient $\mathrm{[-]}$ \\
			$\mu$ & viscosity $\mathrm{[Pa\ s]}$ \\
			$\mu_t$ & turbulent viscosity $\mathrm{[Pa\ s]}$ \\
			$\nu$ & kinematic viscosity $\mathrm{[m^2/s]}$ \\
			$\nu_t$ & turbulent kinematic viscosity $\mathrm{[m^2/s]}$ \\
			$\omega$ & specific dissipation rate $\mathrm{[1/s]}$ \\
			$\sigma_k$ & turbulence closure coefficient $\mathrm{[-]}$ \\
			$\sigma_\omega$ & turbulence closure coefficient $\mathrm{[-]}$ \\
			$\sigma_{\omega 2}$ & turbulence closure coefficient $\mathrm{[-]}$ \\
			$\tau_{ij}$ & subgrid-scale stress tensor
		\end{tabular}
	\end{table}
%
\section*{INTRODUCTION}\label{sec:introduction}
\indentStyle
Turbulent flows constitute most of the fluid flow encountered in many processes. It is characterized by velocity fluctuations in all directions that has an infinite number of degrees of freedom. These flows can be classified as laminar, transition to turbulence and fully developed turbulent flows. The flow in an empty bounded channel becomes laminar until the bulk Reynolds number reaches to 1860 \cite{tsukahara2005dns}. Once the flow speed exceeds that level transition to turbulence occurs, and after a critical Reynolds number, which depends on the flow type, the flow becomes fully-developed turbulence. However, with the presence of an obstruction, simply a cylinder, the fluid separates from both side of the cylinder and two shear layers develop. The formation of these layers causes a phenomenon called von Karman vortex street. On the vortex street, the transition from laminar to turbulent flow occurs at low Reynolds numbers due to the increase in flow instabilities \cite{blevins1990flow}. Numerical simulations of such flows in a channel with high blockage ratio becomes very challenging since the interaction of vortex generated by both cylinder and wall adds more level of complexity. Resolving all scales of eddies in this type of flow with direct numerical simulation (DNS) method becomes infeasible because of the need for significant computational resources. On the other hand, although RANS turbulence models provide slightly better predictions for the transitional to turbulent vortex region, it gives misleading results due to the adverse pressure gradients, flow separation, and laminar behavior in the upstream and further downstream of the cylinder. \\
\indentStyle
The recent development of machine learning based techniques emerges as a promising tool to improve capabilities of computational fluid dynamics (CFD) by integrating data-driven approach to the numerical prediction. In this context, several studies have come to exist to investigate turbulent structures and improve current turbulence models by incorporating the ML techniques more than a decade ago. Milano and Koumoutsakos \cite{milano2002neural} have developed a neural network methodology in order to reconstruct the near wall field in a turbulent flow. They reported that nonlinear neural networks provided better prediction capabilities for the near wall velocity fields. Marusic et al. \cite{marusic2001real} carried out researches on real-time feature extraction of coherent spatiotemporal structures. They successfully extended the existing pattern discovery algorithms to establish the relationship among higher order clusters. These studies initiated a new research area, however, due to limitations in computational power, and lack of data analytics knowledge, not many studies have shown up for a while. \\
\indentStyle
Recently, incorporation of ML techniques in fluid mechanics has gained momentum and applied to turbulence modeling in several different contexts. Generic steps of ML algorithms include a model on training observations and then making predictions on unseen testing observations using the fitted model. Yarlanki et al. \cite{yarlanki2012estimation} used an artificial neural network base ML method to optimize the model constants of the $k$-$\epsilon$ turbulence model. They considered the experimental results of temperature distribution in a data center to be the ground truth training data. The researchers achieved to lowering the RMS error by $25\%$ and absolute average error by $35\%$ compared to the error obtained by using default $k$-$\epsilon$ model constants. Gorle et al. \cite{gorle2012rans} proposed an approach for uncertainty quantification of turbulence mixing models. First, the range of perturbations was obtained as a function of flow features from LES simulations. Then the prediction algorithms were successfully applied to RANS results in order to assess whether the Boussinesq hypothesis is appropriate.\\
\indentStyle
Tracey et al. \cite{tracey2013application} developed an ML algorithm to quantify uncertainties of low-fidelity models by using information from related high-fidelity data sets. Even with the limited data, their method managed to provide upper and lower limits to RANS errors. Duraisamy et al. \cite{duraisamy2015new} inferred the functional form of deficiencies in known closure models by applying inverse problems to experimental data and developed an ML model to obtain more robust and accurate closure models. Ling and Templeton \cite{ling2015evaluation} proposed a classification type ML algorithm to predict the regions in a flow where high RANS uncertainty may occur. They achieved to confident assessment that their method enables the evaluation of RANS uncertainty using data-driven approach and capable of generalizing the markers to flows substantially different from those on which it was trained. Wang et al. \cite{wang2017physics} proposed a data-driven approach to predict Reynolds stress discrepancies in RANS by using regression type ML technique based on Random Forests (RF). They illustrated that the ML algorithms provide noticeable improvements on the baseline RANS simulations at almost no additional computational cost. \\
\indentStyle
In this study, discrepancy of the turbulent kinetic energy (TKE) between $k$-$\omega$ SST RANS and dynamic one-equation LES turbulence models are determined by conducting three-dimensional CFD simulations in a channel containing circular cylinder with high blockage ratio. The simulations are carried out with the Reynolds numbers ranging between $500$ and $1250$ for the training and testing purposes of the RF. The learning algorithm trained on the flow fields at different Reynolds numbers except the one used for testing. Then the trained RF is used to predict the discrepancy of the unseen testing flow. To illustrate the predictive capability of ML, contour plots and profiles obtained at different locations are presented. The results suggest that ML algorithm successfully characterizes the model-based uncertainties in three-dimensional incompressible turbulent flows as a function of features derived from the mean flow properties of RANS.

\section*{MODEL DESCRIPTION}\label{sec:model_desc}

\subsection*{Governing Equations}\label{sec:gov_equ}
In this study, steady and unsteady incompressible turbulent flow is considered for RANS and LES simulations, respectively. The equations governing the flow field are:\\
continuity;
\begin{align}\label{equ: cont}
\frac{\partial u_i}{\partial x_i} = 0
\end{align}
conservation of momentum for steady flow;
\begin{align}\label{equ: steadyRANS}
u_j \frac{\partial u_i}{\partial x_j} = -\frac{1}{\rho} \frac{\partial p}{\partial x_i} + \frac{1}{\rho} \frac{\partial}{\partial x_j} \left( \left( \nu + \nu_t\right) \frac{\partial u_i}{\partial x_j}\right)
\end{align}
conservation of momentum for unsteady flow;
\begin{align}\label{equ: unsteadyLES}
\frac{\partial \bar{u_i}}{\partial t} + \bar{u_j} \frac{\partial \bar{u_i}}{\partial x_j} = -\frac{1}{\rho} \frac{\partial \bar{p}}{\partial x_i} + \nu \frac{\partial^2 \bar{u_i}}{\partial x_j^2} - \frac{\partial \tau_{ij}}{\partial x_j}
\end{align}
\noindent
Here in $\rho$ is the density, $p$ is pressure, $\nu$ is kinematic viscosity, $\nu_t$ is turbulent kinematic viscosity, $u$ is velocity, and $t$ is time. In equation \ref{equ: unsteadyLES} $\bar{u}$ is the filtered velocity obtained by filtering the Navier-Stokes equation and $\tau_{ij}$ is the subgrid-scale stress tensor. \\
\indentStyle
To simulate steady-state turbulent flows, RANS equations are solved along with a two-equation eddy viscosity model proposed by Menter \cite{menter1994two}. In the Menter's $k$-$\omega$ SST model the equations governing the $k$ and $\omega$ are described as:
\begin{equation}\label{equ: menterkEqn}
\begin{split}
\frac{\partial k}{\partial t} + u_j \frac{\partial k}{\partial x_j} = \min\left( \tau_{ij} \frac{\partial u_i}{\partial x_j},10\beta^*k\omega \right) - \beta^*k\omega + \\ \frac{\partial}{\partial x_j}\left[\left( \nu +\sigma_k \nu_t\right) \frac{\partial k}{\partial x_j}\right]
\end{split}
\end{equation} 
\begin{equation}\label{equ: menterOmegaEqn}
\begin{split}
\frac{\partial (\omega)}{\partial t} + u_j\frac{\partial ( \omega)}{\partial x_j} = \frac{\gamma}{\nu_t} \min\left( \tau_{ij} \frac{\partial u_i}{\partial x_j},10\beta^*k\omega \right) - \beta \omega^2 + \\ \frac{\partial}{\partial x_j} \left[ \left( \nu + \sigma_{\omega} \nu_t \right) \frac{\partial \omega}{\partial x_j} \right] + 2(1-F_1) \frac{\sigma_{\omega 2}}{\omega} \frac{\partial k}{\partial x_j} \frac{\partial \omega}{\partial x_j}
\end{split}
\end{equation}
\noindent
Here $\beta^*$, $\sigma_k$, $\gamma$, $\sigma_\omega$, and $\sigma_{\omega 2}$ are closure coefficients and $F_1$ is the blending function which takes different values at the near wall and in the bulk.\\
\indentStyle
To obtain high fidelity results LES simulations are conducted for unsteady flow field using a dynamic one-equation subgrid-scale (SGS) model presented by Kim and Menon \cite{kim1995new}. The dynamic model improves on the limitation of Smagorinsky model by adjusting the proportionality coefficient, $c_v$, defined in the subgrid eddy viscosity, locally during computation instead of defining a priori global constant. The equation governing the $k$ is described as:   
\begin{equation}\label{eqn: oneEddyk}
\frac{\partial k}{\partial t} + \bar{u_i}\frac{\partial k}{\partial x_i} = -\tau_{ij}\frac{\partial \bar{u_i}}{\partial x_j} - \epsilon + \frac{\partial}{\partial x_i} \left( \nu_t \frac{\partial k}{\partial x_i} \right) 
\end{equation}
\noindent
Herein $\tau_{ij}$ is the SGS stress which is defined as a function of turbulent kinematic viscosity, $\nu_t=c_v k^{0.5} \Delta$ where $c_v$ is the model coefficient and $\Delta$ is the grid scale filter. The three terms on the right-hand-side of Eq. (\ref{eqn: oneEddyk}) represent, the production rate, the dissipation rate, and the transport rate of $k$ respectively.

\subsection*{Numerical Model}\label{sec: num_model}
\begin{figure}[t]
	\centering
	\includegraphics{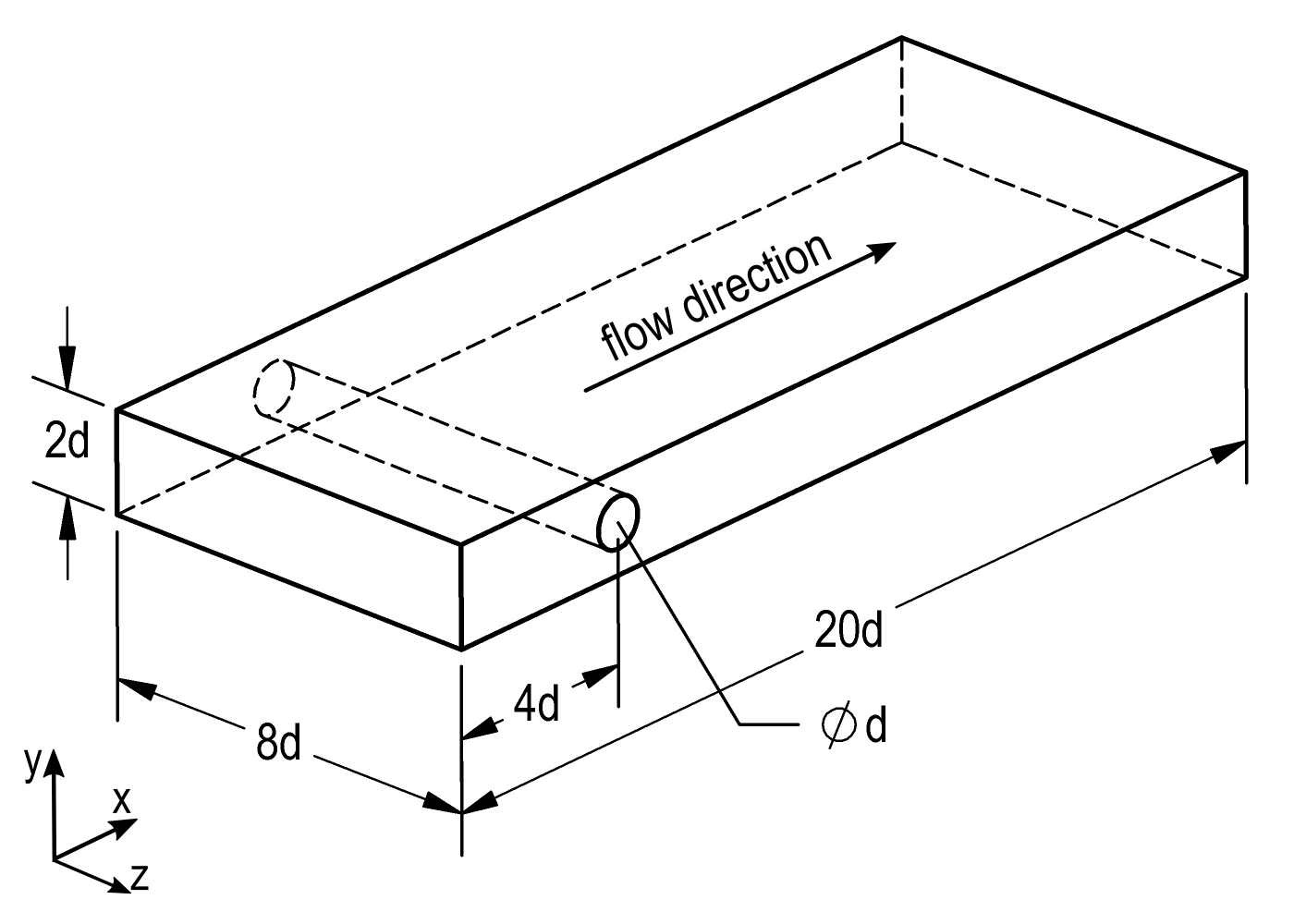}
	\caption{Schematic of the flow domain.}\label{fig:geometry}
\end{figure}

In the current study, the three-dimensional CFD model of flow in a channel containing circular cylinder was developed for steady and unsteady turbulent flows. Schematic of the flow domain along with the flow direction indication is illustrated in Figure \ref{fig:geometry}. Spatially, $x$, $y$, and $z$ directions represent the normalized length, $l=20d$, height, $h=2d$, and width, $w=8d$ of the channel, respectively. The normalization factor is the diameter of the cylinder, $d$ and the blockage ratio of the channel is $0.5$. Reynolds number, $Re=U_a d / \nu$, is defined based on the averaged inlet velocity, $U_a$ and cylinder diameter $d$.\\
\indentStyle
Regarding the boundary conditions, no-slip and no-penetration conditions are applied to the top, bottom, and cylinder walls. To better uncover the three-dimensional flow effects, transitional periodicity is considered in $z$ direction. The intensity of TKE is set to be zero at the inlet since there is no flow disturbance resulting in a transition from laminar to fully developed turbulent flow in the upstream of the cylinder. The dimensionless wall distance, $y^+$ is achieved to be smaller than unity so that no wall functions are used along the walls for TKE.\\
\indentStyle
As for the discretization and solution of the governing equations (Eqs. \ref{equ: cont}-\ref{eqn: oneEddyk}), OpenFOAM-v1706 - an open source finite volume method (FVM) solver - is employed. Specifically, simpleFoam and pimpleFoam algorithms are used for RANS and LES simulations, respectively. To calculate and compare the discrepancies time-averaging of LES results are considered. 

\subsection*{Random Forests}\label{sec: rf}
\begin{figure}[t]
	\centering
	\includegraphics{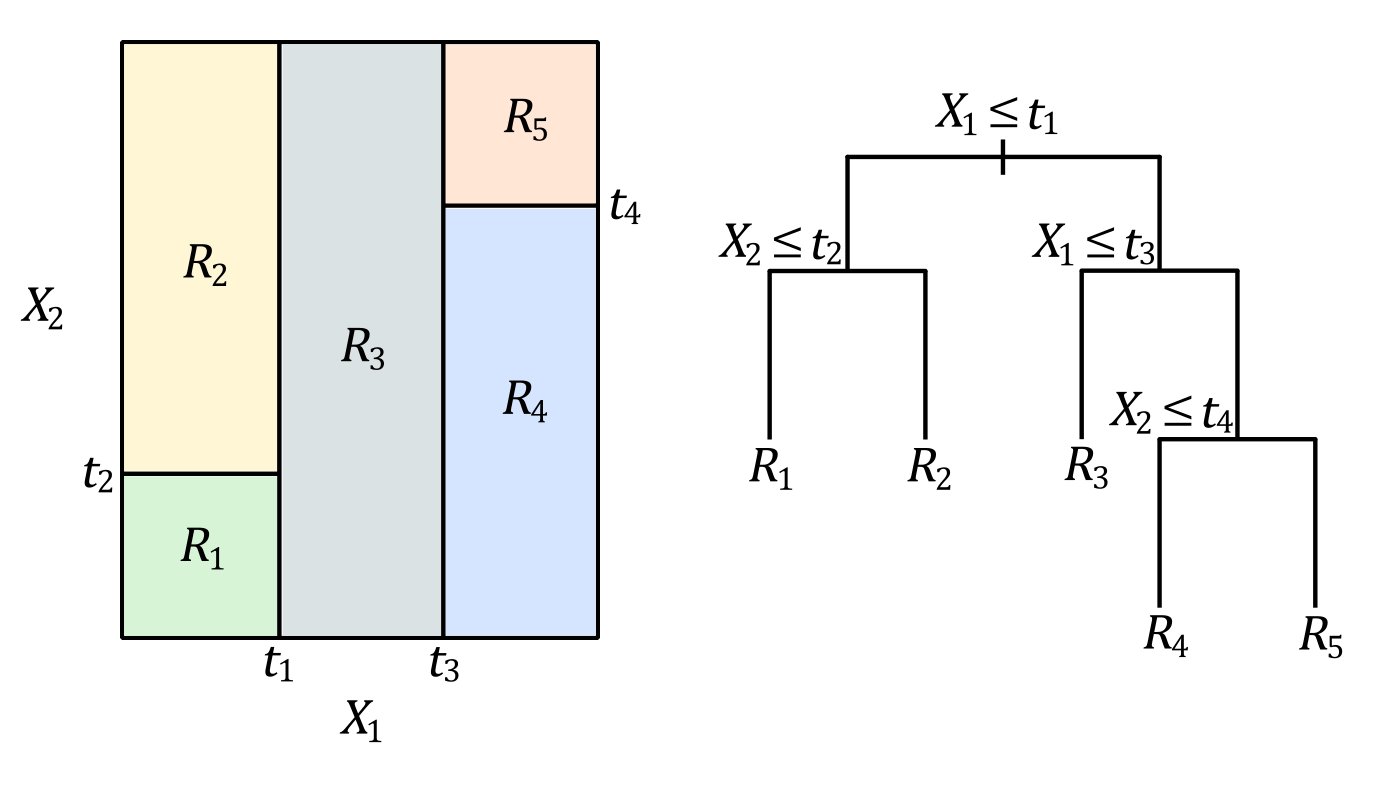}
	\caption{Partition of two-dimensional input feature space into five distinct and non-overlapping regions and its corresponding decision tree representation. $R_k$ is the mean of the observations in region $k$.}\label{fig:dt}
\end{figure}

\indentStyle
Machine learning explores algorithms that can learn from data through fitting a model on training observations and then making predictions for unseen testing observations using the fitted model. Typical data matrix for building a learning algorithm consists of input and output features. The features can be characterized as either numerical or categorical. Examples of numerical features include discrepancy of turbulent kinetic energy between RANS and LES simulations, turbulence intensity, and streamline curvature. Examples of categorical features include a point's uncertainty level in RANS results (low or high) and violation of certain assumptions (yes or no). The problems with categorical output feature are referred to as the classification problems whereas the problems with numerical output feature are referred to as the regression problems. In this study, since our goal is to predict a numerical feature, we focus on the regression problem. \\
\indentStyle
Over the past few decades, a variety of regression techniques have been proposed such as k-nearest neighbors \cite{altman1992introduction}, ridge regression \cite{hoerl1970ridge}, lasso \cite{tibshirani1996regression}, artificial neural networks \cite{yegnanarayana2009artificial}, tree-based methods (e.g., regression trees, random forests, boosting) \cite{james2013introduction}, and support vector regression \cite{drucker1997support}. Among these, we employ the random forests \cite{breiman2001random} in our study since it does not suffer from the curse of dimensionality and provides good predictions with physical interpretations such as the importance of features.\\
\indentStyle
RF is an ensemble of decision trees. Decision trees (DT) divide the input feature space into $K$ distinct and non-overlapping boxes, $R_1, R_2, \ldots, R_K$, with the goal of minimizing the variance within regions given by
\begin{align}
\sum_{k=1}^{K}\sum_{i\in R_k}(y_i - \hat{y}_{R_k})^2,
\end{align}
where $K$ is the total number of regions, $y_i$ is the output feature value of training observation $i$, and $\hat{y}_{R_k}$ is the mean value of the output feature of training observations in region $k$. As for the prediction, when a new observation enters the system, DT uses the mean value of training observations in the region in which the new observation falls. Figure \ref{fig:dt} shows the schematic representation of partition and the corresponding decision tree.\\
\begin{figure}[!]
	\centering
	\includegraphics[scale=0.37]{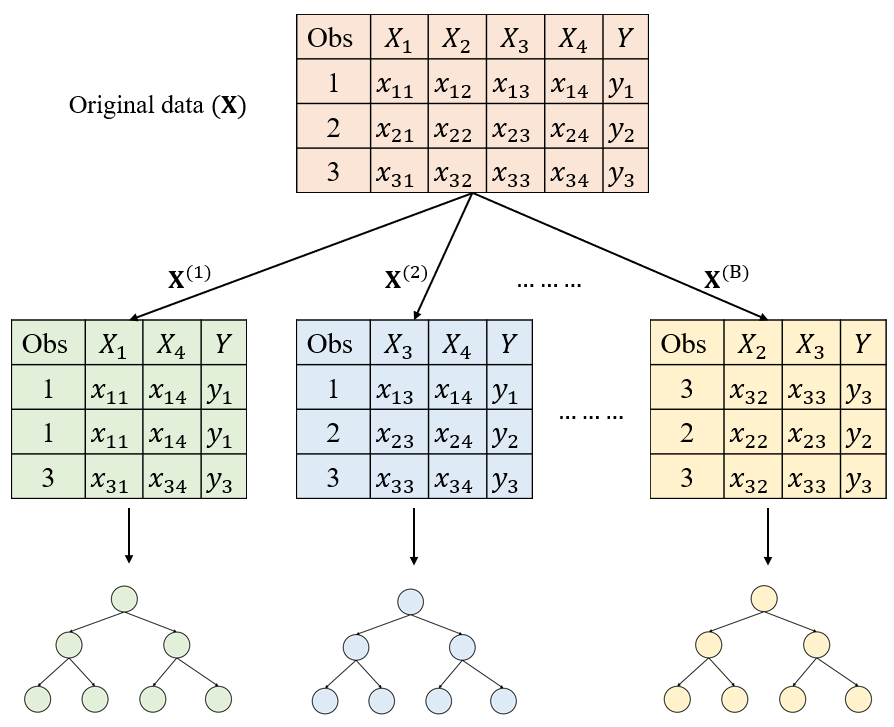}
	\caption{A graphical illustration of random forests algorithm which is the ensemble of $B$ decision trees. Observations are chosen randomly with replacement and two out-of-four attributes are used randomly at each split.}\label{fig:rf}
\end{figure}
\indentStyle
Decision trees are easy to interpret and implement and also computationally inexpensive. However, they are not robust to changes in the training data. Small changes in the training data can lead to large differences in the fitted model and corresponding predictions. Ensemble methods form a "strong learner" using a group of "weak learners." To this extent, RF produces multiple decision trees to address the issue of high variance and then combines them to yield a prediction. First, the training data is split into subsets of observations. The observations and features are chosen randomly at each split. Then a separate decision tree is fitted to each subset. Randomness enhances tree diversity, avoids trees to be very similar to each other, and diminishes the tendency of the model to overfitting. For a given new observation, it is run down all the trees, and the average of the predictions of all trees is considered a single consensus prediction for the new observation. Figure \ref{fig:rf} shows the graphical illustration of randoms forests.\\
\begin{table}[!]
	\small
	\centering
	\caption{Training and testing scenarios.}\label{tab:scenarios}
	\begin{tabular}{C{2cm}L{2cm}L{2cm}}
		\hline
		Scenario No. & Training flows & Testing flow\\
		\hline
		1 & $Re=750$ $Re=1000$ $Re=1250$ & $Re=500$\\
		\hline
		2 & $Re=500$ $Re=1000$ $Re=1250$ & $Re=750$\\
		\hline
		3 & $Re=500$ $Re=750$ $Re=1250$ & $Re=1000$\\
		\hline
		4 & $Re=500$ $Re=750$ $Re=1000$ & $Re=1250$\\
		\hline
	\end{tabular}
\end{table}
\indentStyle
As for the training and testing of the RF algorithm, RANS and LES simulations are carried out using OpenFOAM-v1706 with different flow rates in a three-dimensional transition to turbulent regime, where the data obtained from LES simulations are averaged over time. The mean flow features, which are proposed in \cite{ling2015evaluation,wang2017physics}, are obtained from RANS simulations and used as input and the discrepancy of turbulent kinetic energy is used as the output to the RF algorithm. We use scikit-learn - an open source Python library for ML - for training and prediction purposes of the RF algorithm.
\section*{RESULTS AND DISCUSSION}\label{sec:results}
\begin{figure}[t]
	\centering
	\includegraphics{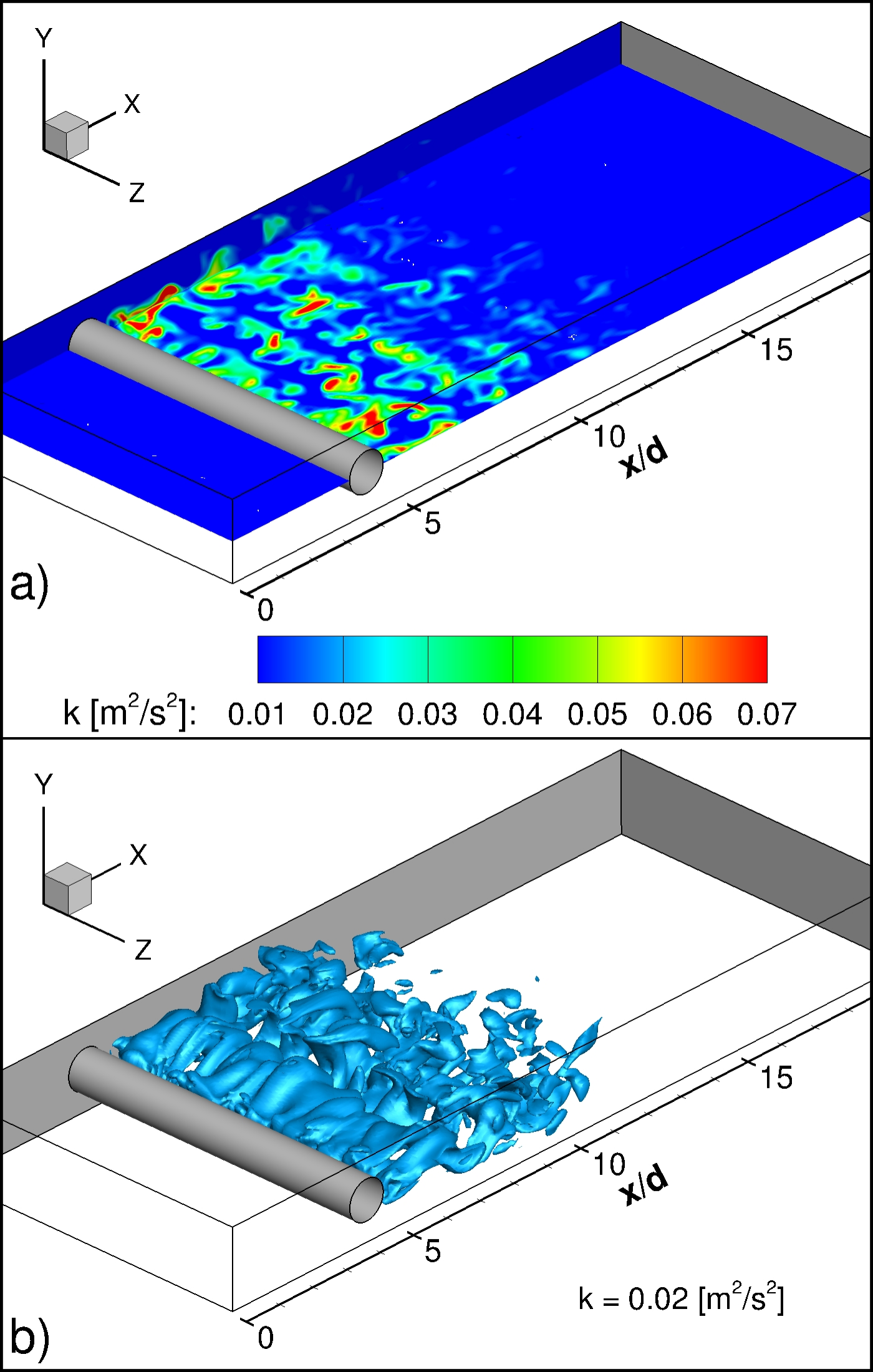}
	\caption{Instantaneous contour (a) and iso-surface (b) of turbulent kinetic energy obtained by LES at $Re = 1250$. Contour is rendered at mid y-plane and iso-surface level is set to be $0.02$.}\label{fig:k_plane_iso_1250}
\end{figure}

\indentStyle
Within the range of Reynolds number considered in this study, the flow remains  laminar in an empty channel. However, submersion of a cylinder interrupting the upcoming stream induce momentum mixing by flow separation and vortex shedding which results in flow instabilities leading to turbulent flow. Figure \ref{fig:k_plane_iso_1250} illustrates an instantaneous contour and iso-surface of TKE for $Re=1250$. As imposed by the inlet condition, TKE intensity remains nearly zero in the upstream flow and transition from laminar to turbulent flow occurs in the downstream. The increased turbulent activity is transported until several diameter away from the cylinder and decays down towards outlet. Such behavior confirms that the flow tends to remain in laminar regime unless it is perturbed. Additionally, the flow in the wake region turns into three-dimensional flow with strong secondary flows, and oscillates which is described by von Karman phenomena, as illustrated in Figure \ref{fig:k_plane_iso_1250}b. The accurate numerical solution of this flow field, essentially, requires a turbulence model which can capture the physics in both upstream and downstream of a cylinder where different flow regimes observed.
\subsection*{RF Prediction of TKE discrepancies at different Reynolds numbers}
\begin{figure*}[h!]
	\centering
	\includegraphics[scale=0.37]{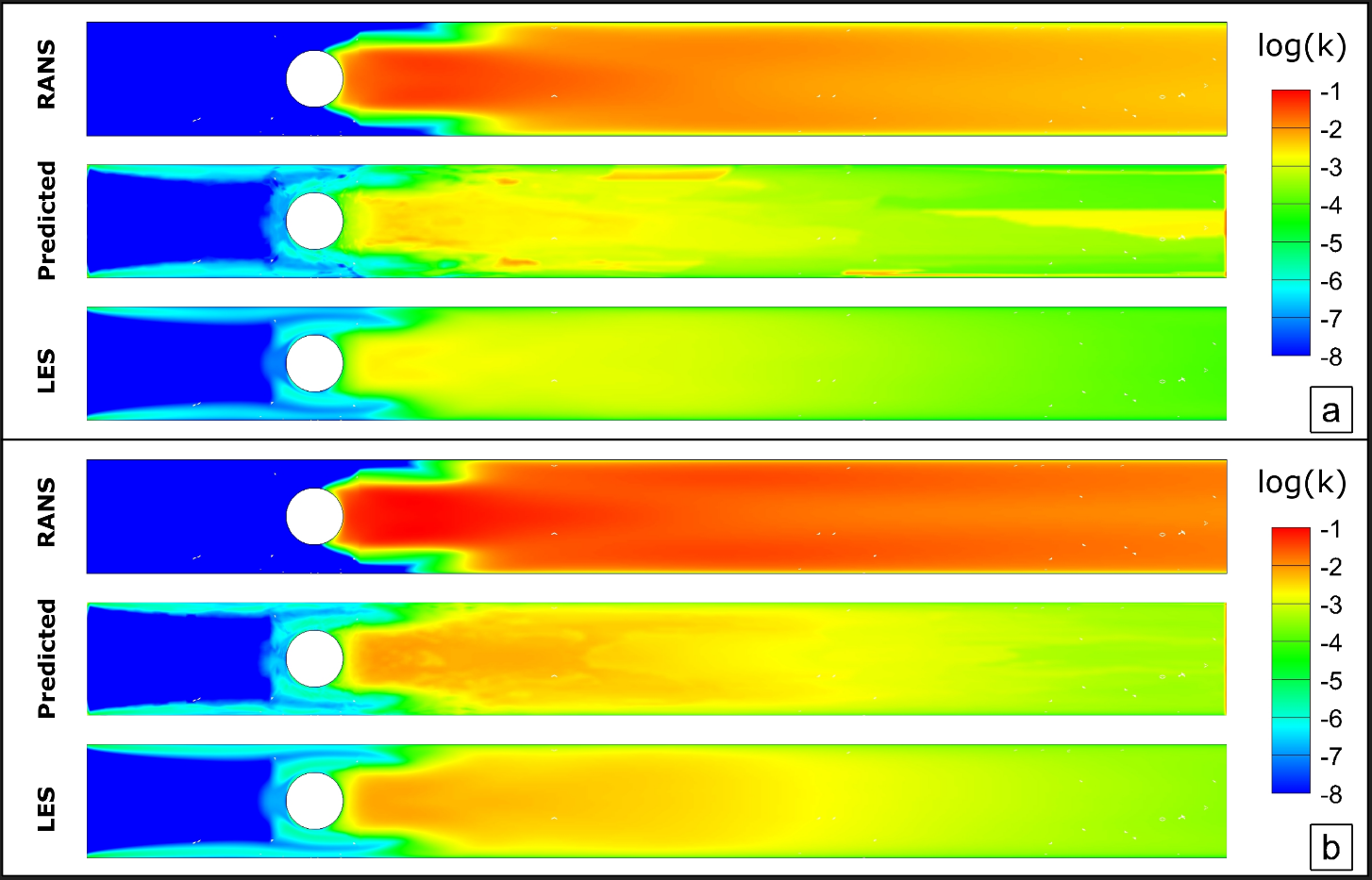}
	\caption{Contours of log turbulent kinetic energy obtained by $k$-$\omega$ SST RANS, RF prediction, and dynamic one-equation LES at (a) $Re = 500$ and (b) $750$. The contours are rendered at mid z-plane.}\label{fig:500_750_kLogMean}
\end{figure*}

\indentStyle
The RANS and LES simulations are carried out with four different Reynold numbers - 500, 750, 1000, 1250 - to obtain the train and test flows. The mean flow features are calculated for each flow using the raw flow properties such as mean pressure, velocity, turbulent kinematic viscosity, and wall distance. Then these features are used as input to the learning algorithm. The input features are normalized so that they are in the range of $[-1,1]$. The log discrepancy of TKE, $\Delta log(k)=log(k)_{LES} - log(k)_{RANS}$, is obtained for each flow as an output. In order to nicely illustrate the TKE discrepancy under significant deviations between RANS and LES simulations, we use the log discrepancy as an output to the learning algorithm. Each flow has 2.8 million data points and 10 input features. As for the training and testing of the RF, we created four different scenarios which are shown in Table \ref{tab:scenarios}. For each scenario, we train and test our model on 8.4 and 2.8 million data points, respectively. We use $B=100$ decision trees to ensemble for each scenario. The value of the number of trees is obtained by 10-fold cross-validation approach \cite{james2013introduction}.\\
\begin{figure}[t]
	\centering
	\includegraphics[scale=1]{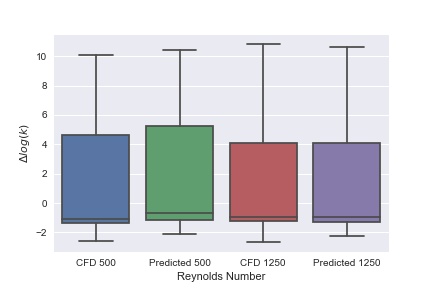}
	\caption{Distributions of log discrepancy of the predicted and true TKE at $Re=500$ and $Re=1250$.}\label{fig:bplot}
\end{figure}
\indentStyle
Figure \ref{fig:500_750_kLogMean} shows contours of log TKE obtained by $k$-$\omega$ SST RANS, RF ML prediction and dynamic one-equation LES at $Re = 500$ and $750$. The contours are depicted at mid z-plane to illustrate the TKE distribution in stream-wise and cross-wise directions. It is observed that RANS fails to capture the elevated TKE inside the boundary layers developing over the walls and cylinder. The near wall behavior of TKE becomes important when it comes to accurate predictions of forces exerted on the cylinder or temperature distribution along the wall as seen in many applications. Moreover, RANS simulations tends to under-predict and over-predict the TKE in the upstream and downstream of the cylinder, respectively, regardless of different flow speeds. Particularly, in the recirculation region right behind the cylinder the severity of over-prediction becomes more pronounced. It is apparent that LES simulations offer better solution at the near wall and cylinder under presence of the adverse pressure gradients and flow separation. Regarding the TKE intensity in further downstream of the cylinder, the dissipation of TKE, as suggested by the time-averaged LES, shows the tendency of transitioning back from the unstable flow to laminar.\\
\begin{figure*}[h!]
	\centering
	\includegraphics[scale=0.37]{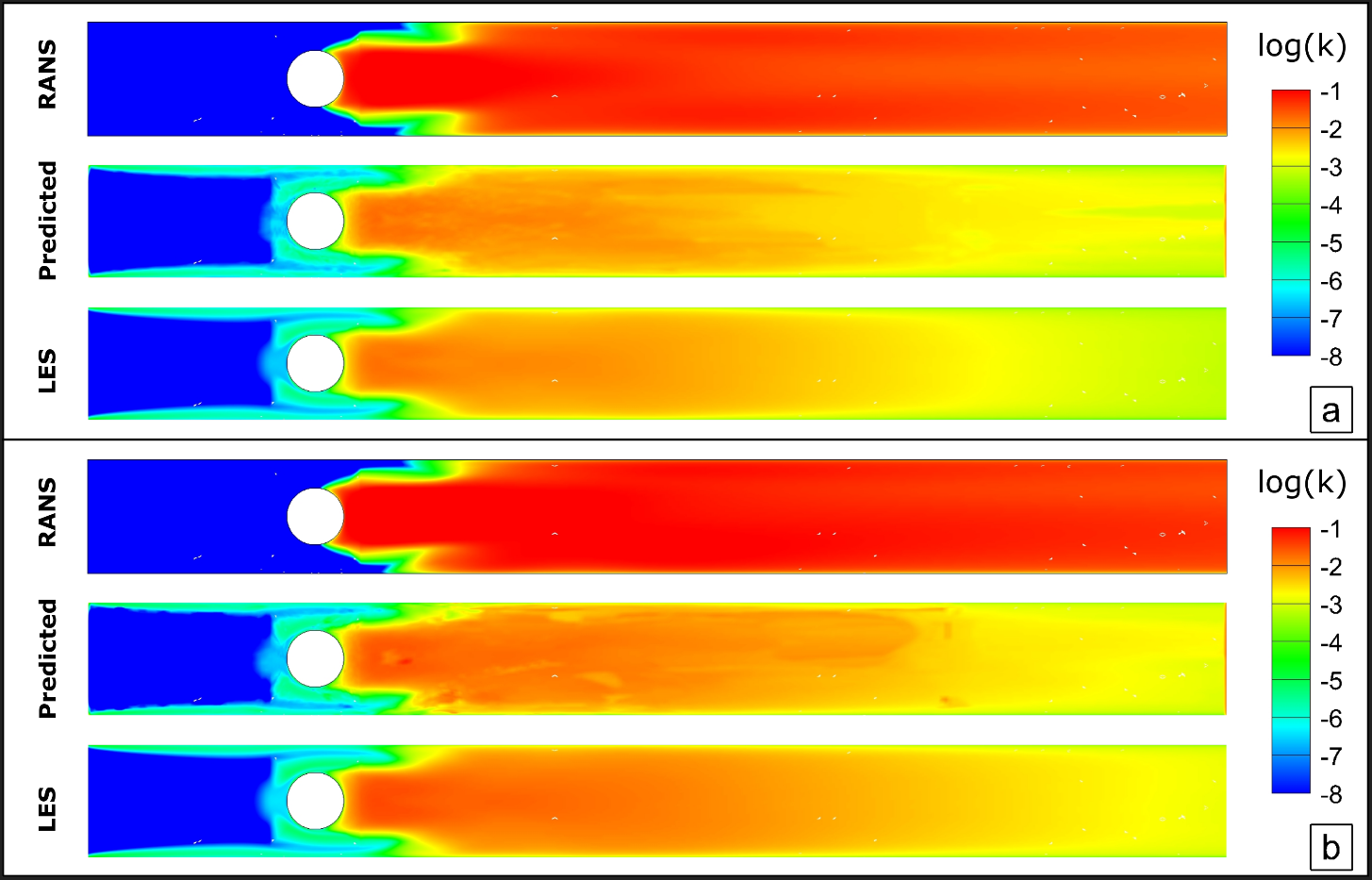}
	\caption{Contours of log turbulent kinetic energy obtained by $k$-$\omega$ SST RANS, RF prediction, and dynamic one-equation LES at (a) $Re = 1000$ and (b) $1250$. The contours are rendered at mid z-plane.}\label{fig:1000_1250_kLogMean}
\end{figure*}
\begin{figure*}[h!]
	\centering
	\includegraphics[scale=0.95]{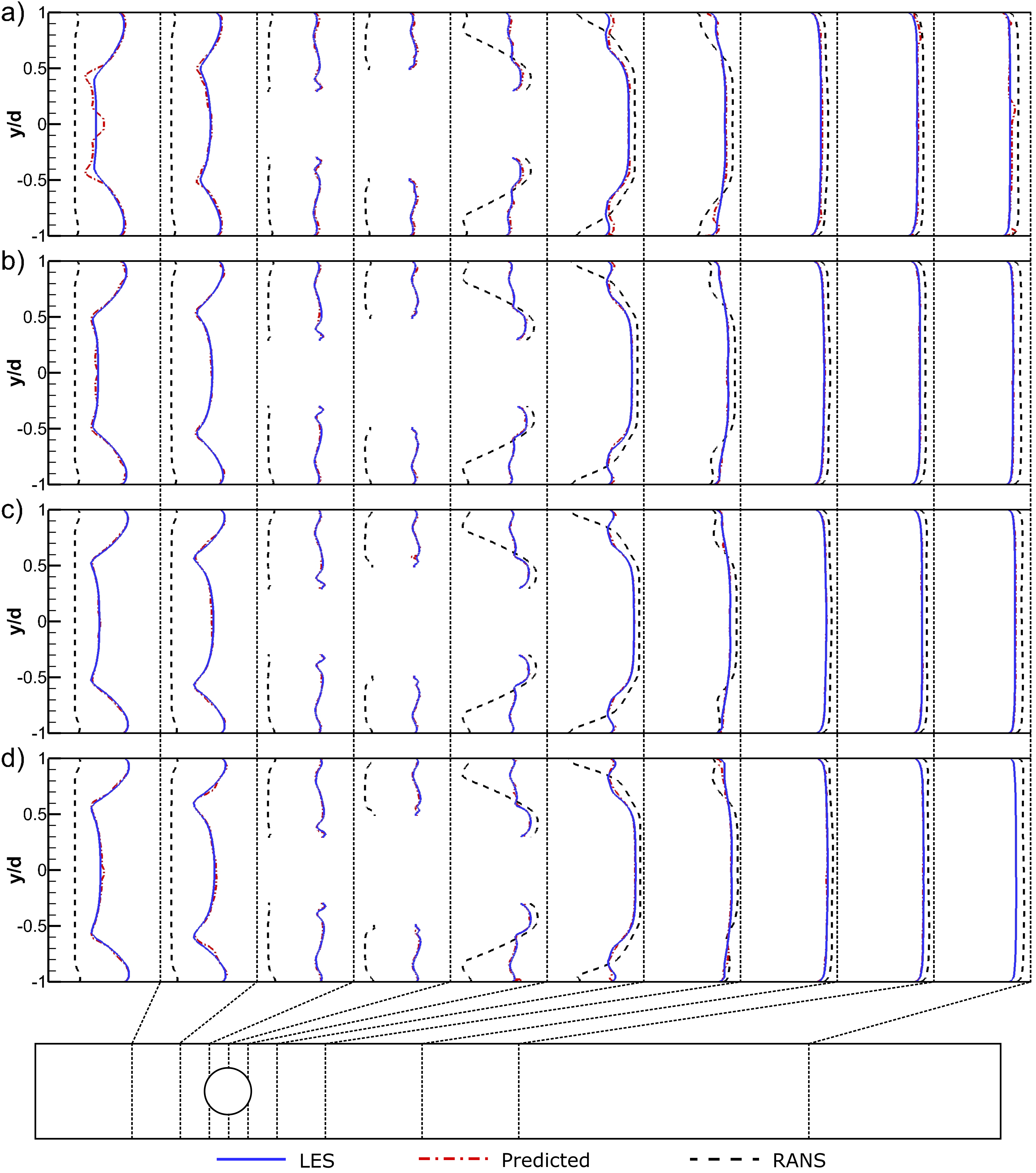}
	\caption{Profiles of log turbulent kinetic energy for RANS, predicted, and LES at (a) $Re = 500$, (b) $Re = 750$, (c) $Re = 1000$, and (d) $Re = 1250$. The profiles are depicted along the cross-wise direction at mid z-plane and 10 different stream-wise locations.  }\label{fig:all_logProfiles}
\end{figure*}
\indentStyle
The RF ML algorithm can address the discrepancies mentioned above and provides satisfactory predictions, as illustrated in Figure \ref{fig:500_750_kLogMean}. It nicely differentiates the over-prediction in the wake region, and under-prediction at near walls and then incorporates the predictive capabilities to improve the results obtained by low fidelity RANS simulations. In particular, the prediction of TKE distribution at $Re = 500$ is sufficient however it deviates more when compared to one at $Re = 750$ since the training is achieved with the flow speeds all greater than $Re = 500$. This indicates the necessity of representative data enough to reveal the flow physics. \\    
%
\indentStyle
Figure \ref{fig:1000_1250_kLogMean} illustrates contours of log TKE at $Re = 1000$ and $1250$. Each subfigure presents the results depicted from mid z-plane for RANS, RF predictions, and LES. As expected, the TKE intensity increases with the increasing flow speed. However, as discussed in the Figure \ref{fig:500_750_kLogMean}, RANS unacceptably over-predicts the intensity in the downstream of the cylinder, particularly within the recirculating region. Apparently, the discrepancy between $k$-$\omega$ SST RANS and dynamic one-equation LES model reaches as much as an order of magnitude. As one of the shortcomings of RANS is discussed in Figure \ref{fig:500_750_kLogMean}, under-prediction of the increase in TKE intensity in the boundary layer persist even at higher Reynolds numbers presented in Figure \ref{fig:1000_1250_kLogMean}.\\
\indentStyle
ML algorithm characterizes the discrepancies and improves on the RANS solutions at both flow speed, as shown in Figure \ref{fig:1000_1250_kLogMean}. In accordance with the issue discussed in Figure \ref{fig:500_750_kLogMean} regarding the training data variability, similar deviation in the prediction of $Re = 1250$ is observed. However, it is realized that the severity of poor prediction is more significant at $Re = 500$. To better assess the performance of prediction between $Re = 500$ and $Re =1250$ - the minimum and the maximum Reynolds number considered in this study - Figure \ref{fig:bplot}, illustrating the distribution of log discrepancy of predicted and true TKE, is presented. Figure \ref{fig:bplot} quantitatively confirms the inference made above in regard to the performance of prediction so that the algorithm predicts $Re = 1250$ better than $500$. Such result is counterintuitive since the prediction of both case is achieved by an algorithm trained in the remaining three flow speeds. As such, it can be inferred that the TKE intensity is not that significant in the lowest Reynolds number for the flow geometry considered in this study. In other words, the flow characteristics at $Re = 500$ fall apart from the rest three flow speeds and some characteristic features of the lowest flow speed is not learned well since such information is not carried by training cases with $Re = 750$, $1000$, and $1250$. This is a clear indication of changing flow phenomena with respect to flow speed in somewhere between $Re = 500$ and $Re = 750$. Beyond predicting the discrepancies, the ML algorithm works as a descriptive analysis tool and reveals crucial information regarding the correlation between TKE intensity and flow speed in transitional flow regimes. \\
\indentStyle
Figure \ref{fig:all_logProfiles} illustrates the profiles of log TKE suggested by RANS, LES, and RF predictions at all Reynolds numbers considered in the present study. The locations of each profile in the stream-wise direction are shown in a geometry schematic at the bottom of Figure \ref{fig:all_logProfiles}. It is worth noting that the profiles appear on left side of y-axis since all TKEs are smaller than unity ending up with a negative value once log of the property is considered. As discussed in Figures \ref{fig:500_750_kLogMean} and \ref{fig:1000_1250_kLogMean}, RANS offers a lower level of TKE in the upstream of the cylinder whereas it predicts higher level in the downstream. The TKE dissipates toward the walls right around and behind the cylinder. However, regarding the LES, the TKE intensity increases in the vicinity of the walls and interacts with the vortex shedding in the wake region resulting in increased turbulent activity at near wall. Such variation in the prediction obtained by RANS and LES proves that both methods behaves considerably different in the transition from laminar to turbulent flow regime. \\
\indentStyle
Profiles depicted from different locations in each flow speed confirms that RF algorithm accurately characterizes the discrepancies and corrects the low fidelity solutions obtained by RANS. In particular, the learning algorithm differentiates the near wall and bulk region so that it captures the phenomena spatially through the flow domain. Such precise prediction offered by the machine learning algorithm also proves the variability of features which provide sufficient information regarding the flow dynamics. 
\section*{CONCLUSION}\label{sec:conclusion}
This paper aims to determine the turbulent kinetic energy discrepancies between RANS and LES and characterize them as a function of features, obtained from mean flow properties of RANS, by means of machine learning algorithms. To accomplish this, three-dimensional CFD simulations in a channel containing a cylinder are conducted for various Reynolds numbers assuring to keep the flow in transition from laminar to the turbulent regime. Then the learning algorithm is trained with different scenarios to predict TKE at different flow speeds.\\
\indentStyle
It is found that (1) RANS and LES predictions significantly deviates, especially in the vicinity of walls and in the downstream of the cylinder; (2) ML successfully predicts the model-based uncertainties in transition to turbulent flow regime; (3) the prediction performance of ML algorithm slightly lower at the lowest Reynolds number due to the fact that the flow characteristic changes at around $Re = 500$; (4) ML approach demonstrates the capability of revealing the correlation between TKE intensity and flow speed in transitional flow regimes.\\
\indentStyle
Overall, the proposed study illustrates how the turbulence models can benefit from the ML algorithms and evidently elucidates that model-based uncertainties of low fidelity approaches can be predicted without requiring a high-fidelity simulation.

\begin{acknowledgment}
	This work used the Extreme Science and Engineering Discovery Environment (XSEDE) for computational need, which is supported by National Science Foundation grant number TG-CTS170051. Specifically, it used the Bridges system at the Pittsburgh Supercomputing Center (PSC). The authors also would like to thank Dr. Alparslan Oztekin for assistance with technical discussion.
\end{acknowledgment}

\bibliographystyle{asmems4}

\bibliography{reference}

\end{document}